\documentclass{jps-cp}
\usepackage{txfonts}

\title{Twisted-Boundary-Condition Formalism for Thermal Transport and an Application to the One-Dimensional XY Spin Chain}

\author{Ryota \textsc{Nakai}$^{1}$, Taozhi \textsc{Guo}$^{2}$ and Shinsei \textsc{Ryu}$^{2}$}

\inst{$^{1}$Department of Physics, Kyushu University, Fukuoka, 819-0395, Japan\\
$^{2}$Department of Physics, Princeton University, Princeton, New Jersey, 08540, USA}

\email{nakai.ryota@mbp.phys.kyushu-u.ac.jp}

\recdate{July 14, 2022}

\abst{
We introduce and formulate the boundary condition twisted by the energy (time translation) in one-dimensional quantum many-body systems. The stiffness against this boundary condition quantifies thermal analogues of the Drude weight and the Meissner stiffness. We apply this formalism to the one-dimensional quantum XY spin chain and estimate the thermal Meissner stiffness.}

\kword{twisted boundary condition, thermal transport, Drude weight, XY model}

\begin{document}
\maketitle

\section{Introduction}

Thermal transport is becoming an indispensable probe of charge-neutral particles such as Majorana fermions \cite{s41586-018-0184-1,s41586-018-0274-0}.
However, thermal transport cannot be analyzed on the same footing as electrical transport, since the driving force, temperature bias, is a statistical force.
The study of the parallelism between these two transport was initiated by Luttinger's work in which a gravitational force was used to effectively represent the temperature bias \cite{PhysRev.135.A1505}, just as the electric potential works for charged particles in the same way as the chemical potential.

In our previous paper \cite{arxiv.2206.00641}, we proposed a new type of parallelism between electrical and thermal transport theory based on the twisted-boundary-condition formalism \cite{PhysRev.133.A171,10.1088/0022-3719/5/8/007,PhysRevLett.39.1167}.
On the electrical conduction side, transport property is categorized into insulators, metals, and superconductors by using two indicators, the Drude weight (charge stiffness) $\bar{D}$ and the Meissner stiffness $D$ \cite{PhysRev.133.A171,PhysRevB.47.7995}.
These indicators measure the stiffness of a quantum system on a ring of perimeter $L$ against the U(1) twisted boundary condition 
\begin{align}
 \psi(x+L,t)=e^{i\phi}\psi(x,t).
 \label{eq:u1twistedboundarycondition}
\end{align}
Here, $\psi$ is the fermion annihilation operator on the spatial and temporal coordinates $(x,t)$, the U(1) phase $\phi$ is identified with the magnetic flux $\Phi$ threaded into the center of the ring via $\phi=e\Phi/\hbar$.
Both stiffnesses are zero in insulators, only the Drude weight is nonzero in metals, and both are nonzero in superconductors.
The Drude weight is identified with the zero-frequency singularity of the ac electric conductivity $\text{Re}[\sigma(\omega)]=2\pi D\delta(\omega)+\sigma_\text{regular}(\omega)$.

 In \cite{arxiv.2206.00641}, we developed the twisted-boundary-condition formalism relevant to thermal transport by introducing what we call ``the energy-twisted boundary condition'' defined by
\begin{align}
 \psi(x+L,t)=\psi(x,t+L\lambda)=e^{i\lambda LH}\psi(x,t)e^{-i\lambda LH}.
 \label{eq:energytwistedboundarycondition}
\end{align}
This boundary condition is intimately related to a thermal transport coefficient, the thermal Meissner stiffness, which will be explained in the next section.
With this formalism, the thermal Meissner stiffness of conformal field theory (CFT), the transverse Ising model and the free fermion model has been studied.

The corresponding bulk transformation is identified with the boost deformation \cite{arxiv.2206.00641}, which is a sort of integrable deformations that keeps the integrals of motion commutable with each other \cite{Bargheer_2009}.
With the boost deformation, two types of the thermal transport coefficients, the thermal Drude weight and thermal Meissner stiffness, can be evaluated.
Applying the boost deformation method to the free fermion model, we studied these stiffnesses by utilizing the solutions of the inviscid Burgers equation.
In particular, the identification of the thermal stiffnesses and the singularity of the ac thermal conductivity has been proved with the free fermion model.
In \cite{arxiv.2206.00641}, we also applied the boost deformation to the Bethe ansatz and studied the linear and nonlinear thermal Drude weight of the XXZ model. 
The nonlinear thermal Drude weight of the XXZ spin chain turned out to agree well with the CFT result unlike the U(1) twist.

In this paper, we apply the energy-twisted boundary condition to study the thermal transport properties of the one-dimensional quantum XY spin chain.
Throughout the paper, we use the unit $\hbar=k_B=1$.

\section{Energy-twisted boundary condition and thermal transport}

Let us review our basic idea of the energy-twisted boundary condition and its relation to the thermal transport coefficients \cite{arxiv.2206.00641}.

\begin{figure}[tbh]
\includegraphics[width=70mm]{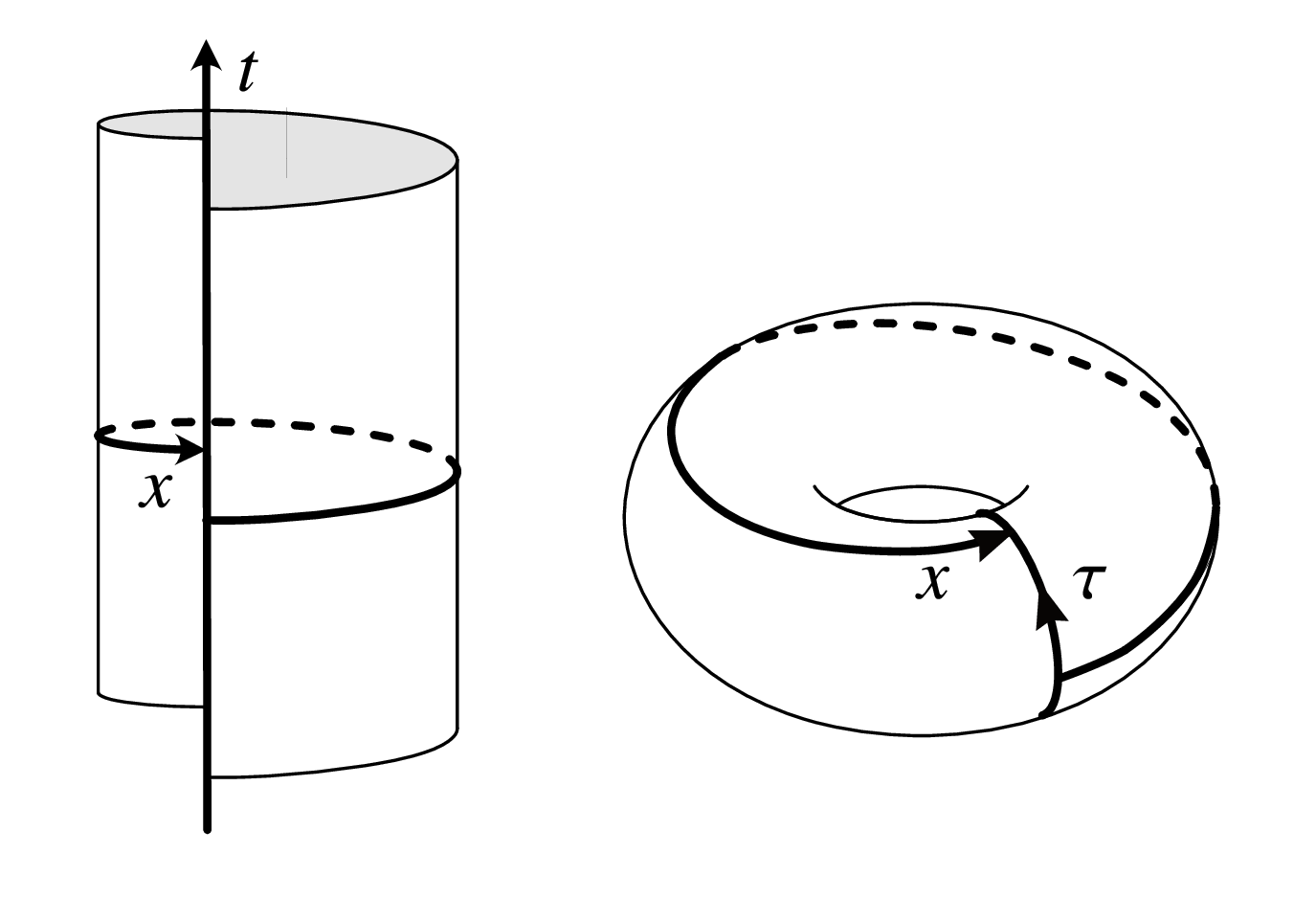}
\caption{Energy-twisted boundary condition of the spacetime cylinder (real time, left) and torus (imaginary time, right).}
\label{twist}
\end{figure}

In (\ref{eq:energytwistedboundarycondition}), the boundary condition is twisted by the Hamiltonian, which is the generator of the time translation.
When the deformation parameter $\lambda$ in (\ref{eq:energytwistedboundarycondition}) is real, the energy-twisted boundary condition is implemented on the spacetime cylinder [Fig.\ref{twist} (left)]. The recipe for the twist is to cut the spacetime cylinder along the $x=0=L$ line, twist one side of the edges along the temporal direction by $\lambda L$, and then glue them to form a cylinder.
When $\lambda$ is pure imaginary, the spacetime is a 2-torus [Fig.\ref{twist} (right)]. The similar process is implemented on the torus. However, in this case the twist is periodic, that is, $\lambda=0$ and $i\beta$ are identified and are related by a modular transformation \cite{Francesco}.

Just as the stiffness against the U(1) phase twist of the boundary condition (\ref{eq:u1twistedboundarycondition}) is the Drude weight and the Meissner stiffness
\begin{align}
 \bar{D} = \frac{L}{2}\sum_{j=1}\frac{e^{-\beta E_n}}{Z}\frac{d^2 E_n}{d\phi^2}\bigg|_{\lambda=0},\quad
 D = \frac{L}{2}\frac{d^2 F}{d\phi^2}\bigg|_{\lambda=0},
\end{align}
the stiffness against the energy-twisted boundary condition (\ref{eq:energytwistedboundarycondition}) is  the thermal Drude weight and thermal Meissner stiffness 
\begin{align}
 \bar{D}^Q = \frac{1}{2L}\sum_{n}\frac{e^{-\beta E_n}}{Z}\frac{d^2 E_n}{d\lambda^2}\bigg|_{\lambda=0},\quad
 D^Q = \frac{1}{2L}\frac{d^2 F}{d\lambda^2}\bigg|_{\lambda=0},
\end{align}
where $E_n$ is an many-body eigenenergy, $Z$ and $F$ are, respectively, the partition function and the free energy.
For the free fermion theory, the thermal Drude weight is identified with the $\omega=0$ singularity of the ac thermal conductivity (for details, see Appendix of \cite{arxiv.2206.00641}).

To implement the energy-twisted boundary condition in quantum many-body systems, we developed both boundary and bulk methods.
In CFT, the energy-twisted boundary condition is a part of standard theory on the torus, where the twist is translated as the modulus.
Away from the criticality where CFT is not available, the energy-twisted boundary condition is incorporated into the transfer matrix method, as we will see in the next section. This method is applicable in any spatial dimensions as long as at least one spatial direction is closed, and thus would be easily applicable to numerical studies.
On the other hand, our bulk method relies on the boost deformation, and so far the equation of the boost deformation is solvable only in one-dimensional integrable models. However, once the deformation of a model is obtained, one can evaluate both thermal stiffnesses.

\section{Transfer matrix method in the XY model}

\begin{figure}[tbh]
\includegraphics[width=80mm]{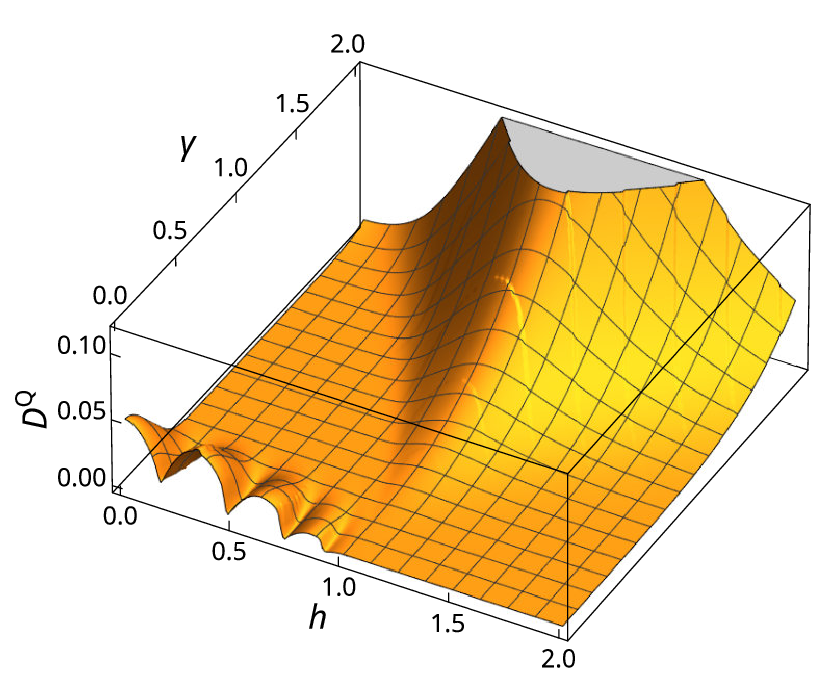}
\includegraphics[width=75mm]{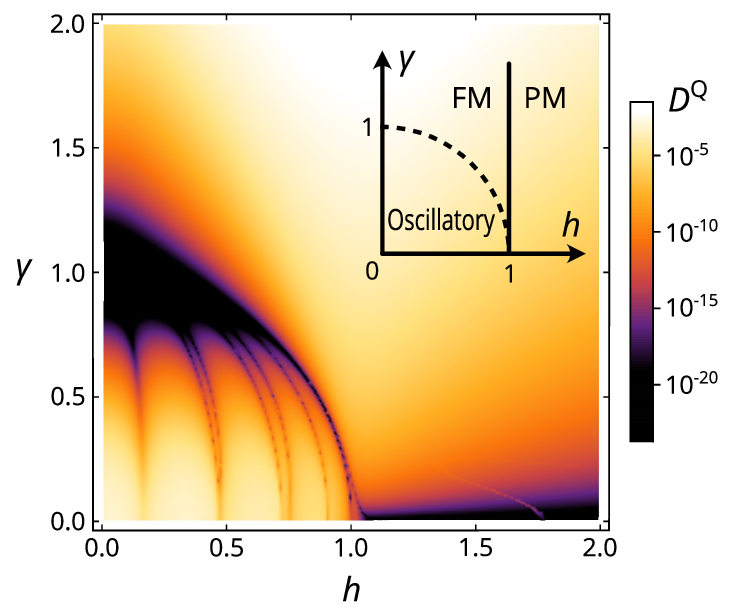}
\caption{The thermal Meissner stiffness of the XY model of length $L=10$ and $J_x+J_y=4$ at inverse temperature $\beta=100$. (Left) on the linear scale and (right) on the log scale. (Inset) the phase diagram of the XY model.}
\label{tms}
\end{figure}

Consider the quantum XY spin chain in one dimension coupled with an external magnetic field 
\begin{align}
 H&=-\sum_{j=1}^L\left(J_xS_j^xS_{j+1}^x+J_yS_j^yS_{j+1}^y-H_z S_j^z\right) \notag\\
 &\equiv
 -\frac{J_x+J_2}{4}\sum_{j=1}^L\left(\frac{1+\gamma}{2}\sigma_j^x\sigma_{j+1}^x+\frac{1-\gamma}{2}\sigma_j^y\sigma_{j+1}^y-h \sigma_j^z\right)
\end{align}
obeying the periodic boundary condition ($S_{N+1}^\alpha=S_1^\alpha$). 
Here, $S^\alpha_j=\sigma_j^\alpha/2$ is the spin $1/2$ operator at site $j$, where $\sigma^\alpha$ is the Pauli matrix, and $\gamma=(J_x-J_y)/(J_x+J_y)$ and $h=2H_z/(J_x+J_y)$ are dimensionless parameters. On $h$-$\gamma$ space, there are three phases [see the inset in Fig.~\ref{tms} (right)]. $h=\pm 1$ corresponds to the Ising-type phase transition between the paramagnetic and ferromagnetic phases. $\gamma=0\,(|h|<1)$ is the isotropic line separating the ferromagnetically ordered phases along $x$ ($\gamma> 0$) and $y$ ($\gamma<0$) directions. On $\gamma^2+h^2=1$, the ground-state wave function is a product of single spin state, while inside of this line ($\gamma^2+h^2<1$) the correlation function is oscillatory.

The partition function is recast on a two-dimensional lattice by the Suzuki-Trotter decomposition $Z=\text{Tr}e^{-\beta H}\simeq \text{Tr}V^M$ \cite{10.1143/PTP.46.1337,10.1143/PTP.56.1454}. By switching from the row-to-row transfer matrix $V$ to the column-to-column one $W$, the partition function is now $Z\simeq \text{Tr}'W^L$, where the trace is over an auxiliary Hilbert space along the Trotter direction. 
The energy-twisted boundary condition is implemented by shifting along the Trotter direction $Z\simeq\text{Tr}'S^aW^L$, where $S$ is a shift operator.
In the XY model, the transfer matrix $W$ is rewritten by the Jordan-Wigner fermion and is block-diagonalized by Fourier modes $e^{i\omega\tau}$ along the imaginary time (Trotter) coordinate $\tau$.
For each block, the shift operator can be replaced by a factor $S^a\to e^{i\omega a}$.

We numerically evaluate the thermal Meissner stiffness of the XY model at sufficiently low temperature by the transfer matrix method (Fig.~\ref{tms}).
The thermal Meissner stiffness vanishes on the edge of the oscillatory phase ($\gamma^2+h^2=1$) as is easily expected from its decoupled ground state.
The Ising critical line ($h=1$) and the isotropic line ($\gamma=0, |h|<1$) correspond to CFTs with the central charge $c=1/2$ and $1$, respectively. The thermal Meissner stiffness derived from CFT at sufficiently low temperature is given by \cite{arxiv.2206.00641}
\begin{align}
 D^Q(T\ll v/L)\simeq \frac{c\pi v^3}{6L^2}.
 \label{eq:thermalmeissnerstiffnesscft}
\end{align}
By the Jordan-Wigner transformation, the transverse XY model is mapped to a free fermion model obeying the Bogoliubov-de Gennes Hamiltonian. The energy dispersion of the Jordan-Wigner fermion is $\epsilon(k)=(J_x+J_y)[(h-\cos k)^2+\gamma^2\sin^2 k]^{1/2}/2$. Thus, the velocity along the Ising criticality ($h=1$) is $v=(J_x+J_y)|\gamma|/2$, and that along the isotropic line ($\gamma=0, |h|<1$) is $v=(J_x+J_y)(1-h^2)^{1/2}/2$.
Substituting these into (\ref{eq:thermalmeissnerstiffnesscft}), the numerical result is explained quantitatively along the Ising criticality. On the other hand, the agreement of (\ref{eq:thermalmeissnerstiffnesscft}) and the numerical result is qualitative on the isotropic line due to the oscillation in a finite-size system.

\section{Conclusion}

We have introduced and formulated the energy-twisted boundary condition as a thermal counterpart of the U(1) twisted boundary condition for electrical transport. The stiffness against this boundary condition is identified with thermal analogue of the Drude weight and Meissner stiffness.
We applied the transfer matrix method to the one-dimensional XY model and estimated the thermal Meissner stiffness.

\section*{Acknowledgments}
R.N.~is supported by JSPS KAKENHI Grant No. JP17K17604 and JST CREST Grant No. JPMJCR18T2.
S.R.~is supported by the National Science Foundation under 
Award No.\ DMR-2001181, and by a Simons Investigator Grant from
the Simons Foundation (Award No.~566116).
This work is supported by the Gordon and Betty Moore Foundation 
through Grant GBMF8685 toward the Princeton theory program.


\begin{thebibliography}{15}
\bibitem{s41586-018-0184-1} M. Banerjee, M. Heiblum, V. Umansky, D. E. Feldman, Y. Oreg, and A. Stern, Nature \textbf{559}, 205 (2018).
\bibitem{s41586-018-0274-0}Y. Kasahara, T. Ohnishi, Y. Mizukami, O. Tanaka, Sixiao Ma, K. Sugii, N. Kurita, H. Tanaka, J. Nasu, Y. Motome, T. Shibauchi, and Y. Matsuda, Nature \textbf{559}, 227 (2018).
\bibitem{PhysRev.135.A1505} J. M. Luttinger, Phys. Rev. \textbf{135}, A1505 (1964).
\bibitem{arxiv.2206.00641} R. Nakai, T. Guo, and S. Ryu, Phys. Rev. B \textbf{106}, 155128 (2022).
\bibitem{PhysRev.133.A171} W. Kohn, Phys. Rev. \textbf{133}, A171 (1964).
\bibitem{10.1088/0022-3719/5/8/007} J. T. Edwards and D. J. Thouless, J. Phys. C: Solid State Phys. \textbf{5}, 807 (1972).
\bibitem{PhysRevLett.39.1167} D. J. Thouless, Phys. Rev. Lett. \textbf{39}, 1167 (1977).
\bibitem{PhysRevB.47.7995}D. J. Scalapino, S. R. White, and S. Zhang, Phys. Rev. B \textbf{47}, 7995 (1993).
\bibitem{Bargheer_2009} T. Bargheer, N. Beisert, and F. Loebbert, J. Phys. A: Math. Theor. \textbf{42}, 285205 (2009).
\bibitem{Francesco}P. Francesco, P. Mathieu, and D. Senechal, Conformal Field Theory (Springer-Verlag, New York, 1997).
\bibitem{10.1143/PTP.46.1337} M. Suzuki, Prog. Theor. Phys. \textbf{46}, 1337 (1971).
\bibitem{10.1143/PTP.56.1454} M. Suzuki, Prog. Theor. Phys. \textbf{56}, 1454 (1976).
\end{thebibliography}
\end{document}